\documentclass[%
 aip,
 amsmath,amssymb,
preprint,%
]{revtex4-1}

\usepackage{graphicx}
\usepackage{dcolumn}
\usepackage{bm}

\usepackage[utf8]{inputenc}
\usepackage[T1]{fontenc}
\usepackage{mathptmx}
\usepackage{subfigure}
\usepackage{etoolbox}
\usepackage{xcolor}

\graphicspath{{Images/}}

\makeatletter
\def\@email#1#2{%
 \endgroup
 \patchcmd{\titleblock@produce}
  {\frontmatter@RRAPformat}
  {\frontmatter@RRAPformat{\produce@RRAP{*#1\href{mailto:#2}{#2}}}\frontmatter@RRAPformat}
  {}{}
}%
\makeatother
\begin{document}


\title{Analytical model of space charge current for a cylindrical porous trap-limited dielectric}
\author{Samra Kanwal}

\author{Chun Yun Kee}%

\author{L. K. Ang}%
\email{ricky\_ang@sutd.edu.sg}
\affiliation{ 
Science, Mathematics and Technology (SMT), Singapore University of Technology and Design, 8 Somapah Road, Singapore 487372
}%

\date{\today}

\begin{abstract}
In this study, analytical models for space charge limited current (SCLC) transport in a porous (or disordered) trap-limited dielectric are derived for both planar and cylindrical configuration.
By considering the porous solid as a fractional object characterized by a parameter $\alpha \leq 1$, we formulate its fractional capacitance and determine the SCLC transport by using the transit time approach. 
At $\alpha$ = 1, it will recover the well-known Mott-Gurney (MG) law and Mark–Helfrich (MH) law for trap-free and trap-limited cases, respectively.
For cylindrical geometry, our findings show an analytical form that is not available from the traditional methods.
We anticipate the proposed analytical model will serve as a useful tool for characterizing the current-voltage measurements in SCLC transport in dielectric breakdown and organic electronics, where spatial porosity of the materials is inevitable.
The introduced fractional parameter $\alpha$ extracted from such characterization can facilitate the quantitative determination of the relationship between spatial porosity and charge mobility.

\end{abstract}
\maketitle
\section{Introduction}
Space charge limited current (SCLC) describes the high current transport in a medium that its electric potential field is strongly influenced by the space charge field of the injected current from the electrode.
If the medium is free space, it is known as the one-dimensional (1D) Child-Langmuir (CL) law \cite{child1911discharge,langmuir1913effect} for a vacuum planar gap of spacing $d$ and biased voltage $V_0$, given by
\begin{align}
    J_{CL}&= \dfrac{4 \epsilon_0}{9} \sqrt{\dfrac{2 e}{m}} \dfrac{V_0^{3/2}}{d^2},
\end{align}
 where $\epsilon_0$ is free-space permittivity, $e$ and $m$ are the electron's charge and mass.
 Modern revision of the 1D CL law includes various effects such as
 quantum CL law, \cite{lau1991quantum,ang2003new,ang2004simple,bhattacharjee2008experimental}, multi-dimensional models \cite{luginsland1996two,lau2001simple,koh2005three,PRL2001edgeCL,AL2021PHP}, short-pulse effects \cite{valfells2002effects,ang2007ultrashort,pedersen2010space}, Coulomb blockade \cite{ZhuAPL2011}, electromagnetic effects \cite{chen2011two}, rough cathode \cite{zubair2016fractional}, sharp tip \cite{ZhuPHP2015}, non-uniform emissions \cite{LAU2020TED,LauTPS2021,sitek2021space,JH2022PRApp,AV2021PRApp} and others summarized in recent papers \cite{APR2017,JAP2021review}.
For a trap-free solid, its SCLC model is known as the 1D Mott-Gurney (MG) law \cite{lampert1970current}:
\begin{align}
    J_{MG}&= \dfrac{9 }{8}\epsilon \mu \dfrac{V_0^{2}}{d^3},
\end{align}
where $\epsilon$ is the permittivity of the solid, and $\mu$ is the mobility. 
For a trap-limited solid with exponentially distributed traps, it is given by the Mark-Helfrich (MH) law \cite{mark1962space,lampert1970current}:
\begin{align}
J_{MH}&= N_c \mu e^{1-l} \Big [\dfrac{\epsilon l}{N_t (l+1)}  \Big]^l \left( \dfrac{2l+1}{l+1} \right)^{l+1} \dfrac{ V_0^{l+1}}{d^{2l+1} }.
\end{align}
Here, $N_c$ denotes the effective density of states corresponding to the energy at the bottom of the conduction band, $N_t$ denotes the total trapped electron density, and $l=T_t/T \ge 1$ is the ratio of distribution of traps to the free carriers. Note the MH law will recover the same voltage scaling of the MG law at $l = 1$.  
These MG and MH laws have been extended to higher dimensional models \cite{chandra2007two,kee2022analytical}, transition models \cite{Chandra_2009,JAP2009,PRL2012,2012organic,2015organic,ZhuTPS2021}, quantum MG law \cite{JAP2015QMG}, 2D materials \cite{greenwood2016relativistic,ang2017relativistic}, arbitrary trap distribution model \cite{stockmann2022exact} and atypical geometries \cite{JAP2021review}.
Recent applications have rekindled interests of SCLC in solids, such as charge transfer \cite{rohr2018charge,fratini2020charge}, Schottky emission \cite{akgul2021investigation}, doped nanowires \cite{vermeersch2022dx}, light-emitting diodes \cite{torricelli2010space,park2018observation}, organic electronics \cite{carbone2005space,carbone2009trapping,craciun2008universal,matsushima2009observation,rojek2019ultra}, and silicon Schottky junctions loaded with nanocrystallines \cite{tsormpatzoglou2006deep}.

Unlike inorganic materials, spatial disorders or imperfections within organic materials cannot be ignored. 
Thus one may wonder if the traditional MG and MH law are accurate for the characterization of its SCL current transport in a spatially disordered solid. 
Such characterization in fitting the current-voltage (I-V) measurement with a correct model is important to estimate the mobility of charge transport.
In a recent paper based on the fractional dimension, a model \cite{zubair2018thickness} was developed to calculate the SCLC for a spatially disordered planar solid, which has shown better agreements over a wide range of organic materials.
Its formulation is however limited to a simple planar geometry that cylindrical trap-free or trap limited diode with spatial disorders has not been developed.
Such development is not trivial that even for the relatively simple cylindrical SCLC model for a vacuum gap remains analytically unsolved
\cite{PRL2013,AL2019APL,APL2021-1,APL2021-2,AL2021PHP,ALTPS2022}.

Thus, we are interested to obtain the analytical results of both MG and MH laws in a cylindrical diode for a trap-limited dielectric.  
By considering the porous dielectric as a fractional object characterized by a parameter $\alpha \leq 1$ (a perfect solid is defined at $\alpha$ = 1), we first solve analytically for its cylindrical capacitance as a function of $\alpha$, and apply the transit-time method \cite{PRL2013,zhu2011analytical} to calculate the corresponding SCLC.
In doing so, we need to extend our recent fractional and planar capacitance model \cite{kanwal2022capacitance} to a cylindrical trap-limited solid. 
From the obtained capacitance results, we will derive the analytical trap-limited SCLC model for arbitrary $\alpha \leq 1$ over a wide range of parameters.
For verification, our model will recover to the well-known classical SCLC models at $\alpha$ = 1.

 Note the fractional calculus \cite{TARASOV2008,2019book} have been widely used to model a wide range of topics that is not possible to describe in details.
 Some examples are hydrodynamics \cite{TARASOV2005, balankin2012map}, thermodynamics \cite{tarasov2016heat}, electrodynamics \cite{zubair2012differential}, chaotic systems \cite{bukhari2022design}, quantum transport \cite{xu2021quantum,kotimaki2013fractal}, exciton binding energy \cite{MZ2020PRApp}, tunneling \cite{MZ2018TED}, Fresnel coefficient \cite{MZ2018JAP}, growth dynamics \cite{Chaos2022,Safdari2016}, and others \cite{Sun2018}. 
 In Sec. II, we first illustrate our methods in re-deriving the prior results \cite{zubair2018thickness,zhu2011analytical} using this fractional capacitance approach \cite{kanwal2022capacitance} to prove its consistency.
In Sec. III, we present the new findings of the fractional capacitance for a cylindrical diode and its corresponding trap-free and trap-limited SCLC line density.
Finally, the paper concludes with a summary and some future works. 
The formulation presented in this paper can be regarded as the fractional models of the MG and MH laws for both planar and cylindrical geometries shown in Table I.
Prior results and other methods are also added in the table for comparison.
\begingroup
\setlength{\tabcolsep}{8pt}
\begin{table*}[ht]
\centering
\caption{Space charge limited current (SCLC) models of trap-free (MG law) and trap-limited solid (MH law) for planar and cylindrical diode}
\label{table-Trap_cases}
\begin{tabular}{cccccc}
\hline \hline
{}&\multicolumn{1}{l}\centering{\bf{Planar}} & {} &{}  &\multicolumn{1}{l}\centering{\bf{Cylindrical}}  &{} \\          \cline{2-3}  \cline{5-6}                                                             
\textbf{Traditional approach} & Non-fractional & Fractional &{} &  Non-fractional & Fractional\\
\hline
Trap-free solid & MG law\cite{mott1948electronic}  & Zubair \cite{zubair2018thickness} &{} & Meltzer\cite{meltzer1960space} &
\\
Trap-limited solid & MH law \cite{mark1962space} & Zubair \cite{zubair2018thickness} &{} &  & 
\\
\hline
\textbf{Capacitance approach} 
\\
\hline
Trap-free solid & Zhu\cite{zhu2011analytical} & Eq. (\ref{eq:frac_MG}) &{} & Zhu \cite{zhu2011analytical} & Eq. (\ref{eq:frac_cyl_MG})
\\
Trap-limited solid &  Zhu\cite{zhu2011analytical}& Eq. (\ref{eq:frac_MH}) &{} & Zhu \cite{zhu2011analytical} & Eq. (\ref{eq:frac_cyl_MH})

             \\ \hline \hline
\end{tabular}
\end{table*}
\endgroup

\section{Fractional Planar SCLC Model}

In this section, we first show the derivation of the 1D MG law and 1D MH law for a porous
dielectric by combining the concept of fractional capacitance \cite{kanwal2022capacitance} and transit-time method \cite{zhu2011analytical} in a
planar geometry. The objective is to show the consistency of fractional modeling in recovering the
same analytical results based on prior traditional 
approaches.

\subsection{Planar MG law for a trap-free solid}
Consider a porous dielectric sandwiched between two electrodes at $z$ = 0 and $z = d$ with a biased voltage of $V_0$. 
The 1D dielectric planar slab has an area dimension of $L >> d$ at the $x-y$ plane. 
For simplicity, let's assume a one-dimensional model in that we only have porosity in the $z$ direction, where the porosity is characterized by $0 < \alpha \leq 1$.
At $\alpha$ = 1, the model describes a perfect non-porous solid.
Without any traps, the capacitance of such a porous dielectric is determined by solving the Laplace's equation in non-integer dimensional spaces \cite{kanwal2022capacitance}, which gives
\begin{equation}\label{eq:frac-capacitance}
{C} = \frac{{Q}}{{V_0}}= \left(\frac{\epsilon_0 \epsilon_r}{\pi^{\alpha/2}}\right)\Gamma\left(\frac{\alpha}{2}\right) \left(\frac{L^2}{d^\alpha} \right),
\end{equation}
where $\Gamma$ is the gamma function and the total charge $Q$ is

\begin{equation}
{Q} = 
\left(\frac{\epsilon_0 \epsilon_r}{\pi^{\alpha/2}}\right)\Gamma\left(\frac{\alpha}{2}\right) \left(\frac{L^2}{d^\alpha} \right) V_0.
\end{equation}
The velocity $v$ of the charge transport inside the dielectric is determined by the mobility $\mu$ and electric field $E$ in the z-direction given by $v = \mu \times E$. The electrostatic potential $V(z)$ without the space charge effects can be solved by using the fractional Laplace's equation given by \cite{zubair2016fractional}
\begin{equation}
 \nabla_\alpha^2 V(z) = \frac{1}{c^2(\alpha,z)}  \left(\frac{\partial^2V(z)}{\partial z^2}- \frac{\alpha-1}{z}\frac{\partial V(z)}{\partial z} \right)=0, \,\, c(\alpha,z)=\frac{\pi^{\alpha/2}}{\Gamma(\alpha/2)}z^{\alpha-1}.
\end{equation}
The solutions of the electric field 
$E = |-\nabla_{\alpha }V(z)|$ and the electric potential $V(z)$ are 

\begin{equation}
E =\frac{\alpha \Gamma(\alpha/2) V_0}{\pi^{\alpha/2} {{d}^\alpha}}, 
\end{equation}
\begin{equation}
V(z)=\frac{V_0}{d^\alpha}z^\alpha.
\end{equation}

For electrons transported from the cathode to the anode, its $z$-direction transit time across the porous solid is  
\begin{align}
\tau &= \dfrac{\int_0^d d z^\alpha}{\nu} = (\frac{\pi^{\alpha/2}{\int_0^d{z}^{\alpha-1}dz}}{ \Gamma(\alpha/2)} )\frac{1}{\nu}\end{align}
where
$d z^\alpha = \dfrac{\pi^{\alpha/2}}{\Gamma(\alpha/2)}z^{\alpha-1} dz$ is a differential fractional length.
Based on $v = \mu \times E$ and $E$ from Eq. (7), we have
\begin{align}
\tau &= \frac{\pi^{\alpha/2}{{d}^{\alpha}}}{\alpha \Gamma(\alpha/2)}  \frac{ \pi^{\alpha/2}{{d}^{\alpha}}V_0^{-1}}{\mu \alpha \Gamma(\alpha/2)}. 
\end{align}
The transported current density scaling is $J \propto \frac{Q/A}{\tau}$, and the emitting area A is $L^2$.
By using Eq. (5) and (9), we obtain
\begin{align}
J &= K \times \frac{\epsilon \mu \alpha^3 \Gamma(\alpha/2)^3 V_0^{2}}{\pi^{3\alpha/2} {{d}^{3\alpha}} }.
\end{align}
Here, the numerical constant $K = 9/8$ may be determined as follows.
Based on the relation of $J = \rho \times v$, and $v = \mu \times E$, Eq. (11) indicates $V(z)$ has a form of
\begin{equation}
V(z) = \frac{V_0}{ {{d}^{3\alpha/2}}  } {{z}^{3 \alpha/2}},
\end{equation}
which can be used to solve for the arriving velocity $v$ and the charge density $\rho$ at $z = d$:
\begin{align}
\nu(d) &= \mu \nabla_\alpha V(d) =\frac{3}{2}\mu \frac{\alpha \Gamma(\alpha/2) V_0}{\pi^{\alpha/2} {{d}^\alpha} },
\end{align}
\begin{align}
\rho(d) &= \epsilon \nabla_{\alpha}^{2} V(d)= \frac{3}{4}\epsilon \alpha^2 \frac{\Gamma(\alpha/2)^2}{\pi^{\alpha} {{d}^{2 \alpha}   } }  V_0.
\end{align}
Thus based on $J_{MG} = \rho({{d}}) \nu({{d}})$, the fractional MG law for a porous trap-free dielectric is
\begin{align} \label{eq:frac_MG}
J_{MG}&= \frac{9}{8} \epsilon \mu \alpha^3 \Big[\frac{\Gamma(\alpha/2)}{\pi^{\alpha/2}} \Big]^3 \frac{V_0^{2}}{  {{d}^{3 \alpha} } }.
\end{align} 
At $\alpha = 1$, it recovers to Eq. (2), which is the classical 1D MG law for a $perfect$ trap-free solid without any porosity effect.

\subsection{Planar MH law for a trap-limited solid}
By using the same capacitance model described in Eq. (4), we re-derive the 1D planar MH law for a trap-limited porous dielectric slab. 
It is assumed that the slab is filled with an exponentially distributed trap in energy, which can be defined by the density of trap states as a function of band energy $E$: $N(E)= N_0 e^{(E-E_c)/k T_t}$, where $E_c$ is energy at the bottom of the conduction band, $T_t$ is the temperature of the distribution, $k$ is the Boltzmann's constant and $N_t$ is the total electron trap density. 
When the solid is biased under an electric field, we assume that the quasi-Fermi level energy is likewise constant along the z-direction, denoted by $E_f$. 
Based on this condition, the density of the trapped electrons is
\begin{align}
n_t &= \rho_t/e = N_t e^{(E_f-E_c)/k T_t},    
\end{align}
 and the density of free electrons at the valence band is,
 \begin{align}
 n_f = \rho_f/e = N_c e^{(E_f-E_c)/k T},    
 \end{align}
 where $N_c$ is the effective density of states at the valence band, $\rho_t$ and $\rho_f$ are the charge densities of trapped and free electrons. 
By eliminating $E_f$ from Eqs. (16) and (17), we can obtain the relationship between $n_f$ and $n_t$ as
\begin{align}
 n_f &= N_c \left( \frac{n_t}{N_t} \right)^l.   
\end{align}
Taking into account that the bound charges on the capacitor are the injected electrons causing the traps to be filled up, we have 
$n_t = {C} \times V_0 / (e \int_0^d d z^\alpha A)$, and $\int_0^d d z^\alpha= \frac{\pi^{\alpha/2}}{\Gamma(\alpha/2)}\frac{{{d}^\alpha}}{\alpha}$,
which gives
\begin{equation}
n_t = \frac{\epsilon \alpha^2 \Gamma(\alpha/2)^2 V_0}{e \pi^\alpha{{d}^{2\alpha}} }.
\end{equation}
Combining Eqs. (18) and (19), the density of free electrons is
\begin{align}
\rho_f&= 
e N_c \left( \frac{\epsilon \alpha^2 \Gamma(\alpha/2)^2 V_0} {e \pi^\alpha {{d}^{2\alpha}} N_t } \right)^l.
\end{align}
Based on the same approach above in determining the fractional capacitance in Eq. (4) and mobility condition, the trap-limited SCL current density $J = \rho_f \times \nu $ becomes
\begin{align}
J = \frac{e^{1-l}N_c \alpha^{2l+1} \Gamma(\alpha/2)^{2l+1}\mu V_0^{l+1} }{ N_t^l \pi^{\alpha/2+\alpha l} {{d}^{2\alpha l +\alpha}} }.
\end{align}
From the continuity equation ($J$ is a constant), we have 
$J \propto (V^{l+1}(z))/( {{z}^{2\alpha l +\alpha}}  )
= (V_0^{l+1})/( {{d}^{2\alpha l +\alpha}} )$, and this implies that the electrical potential and electric field are in the form of
\begin{align}
 V(z)&= \frac{V_0}{ {{d}^{\frac{2\alpha l +\alpha}{l+1}}}  }  {{z}^{\frac{2\alpha l +\alpha}{l+1}}} , \\
 \label{eq:frac_pl_trap_E}
 E &= |- \nabla_\alpha V(z)| =\Big[\frac{2l \alpha+\alpha}{l+1}\Big]\frac{\Gamma(\alpha/2)}{\pi^{\alpha/2}} \frac{V_0}{{{d}^{\alpha }}}.
\end{align}
Using Poisson’s equation in fractional dimensions,  trapped charge density can be calculated as $\rho_t =  \epsilon \nabla_{\alpha}^2 V(z)$ expressed as
\begin{align}
\rho_t(z) = \frac{\epsilon \alpha^2 l(2l+1) \Gamma(\alpha/2)^2}{(1+l)^2 (\pi^{\alpha/2})^2} \frac{V_0}{ {{z}^{2\alpha }}},
\end{align}
and the free electron density is
\begin{align}
\label{eq:frac_pl_trap_rf}
\rho_f(z) = e N_c \Big[\frac{\epsilon \alpha^2 l(2l+1) \Gamma(\alpha/2)^2}{(1+l)^2 (\pi^{\alpha/2})^2 e N_t} \frac{V_0}{ {{z}^{2\alpha }}} \Big]^l.
\end{align}

By evaluating the electron's velocity [$v=\mu E$, Eq. (\ref{eq:frac_pl_trap_E})] and free charge density [Eq. (\ref{eq:frac_pl_trap_rf})] at the anode ($z=d$)  and applying $J_{MH} = \rho_f (z=d) \times \nu (z=d)$, we obtain the 1D fractional trap-limited current density or MH law: 
\begin{align} \label{eq:frac_MH}
J_{MH}
&= N_c e^{1-l} \mu \Big[\frac{\Gamma(\alpha/2)}{\pi^{\alpha/2}}\Big]^{2l+1} \Big[\frac{\epsilon \alpha l }{N_t (1+l)}\Big]^l \left(\frac{2l\alpha+ \alpha}{1+l} \right)^{1+l} \frac{V_0^{1+l}}{{{d}^{2\alpha l+\alpha }}}.
\end{align}
At $\alpha = 1$, it recovers to Eq. (3), which is the classical MH law for a $perfect$ trap-limited solid without any porosity effect.
It is important that while Eqs. (15) and (26) are identical to the prior results reported \cite{zubair2018thickness} as shown in Table I, the derivation approach is different in this paper, which is based on capacitance and transit time.
Note these planar fractional models were shown to have good agreements in its comparison with various experimental measurements \cite{zubair2018thickness}.

\section{Fractional cylindrical SCLC Model}
In this section, we show the new analytical cylindrical models of MG law and MH law for a
porous dielectric cylindrically shaped capacitor.
Consider that the anode (outer electrode) and the cathode are located at $r=b$ and $r=a$ with $b>a$, where $r$ denotes the radial coordinate.
For simplicity, $r$ is normalized by $a$ as $\bar{r}=r/a$ and electrical potential is normalized by $V_0$ as $\bar{V}= V/V_0$. 
The bar over the parameters represents the normalized expressions. 
The porosity of the cylindrical dielectric (in $r$-direction) is characterized by $0 < \alpha \le 1$.
By converting the fractional Laplacian and gradient operators from Cartesian to cylindrical coordinate
\cite{zubair2023coordinate}, the fractional cylindrical capacitance model can be determined analytically as follows:
\begin{align}
\begin{split}
\nabla_\alpha^2\bar{V}&= \frac{1}{c^2(\alpha,\bar{r})}\left( \frac{\partial \bar{V}^2}{\partial \bar{r}^2}-\frac{1}{\bar{r}}\frac{q}{p}\frac{\partial \bar{V}}{\partial\bar{r}}      \right)   = 0  ,\,\ c(\alpha,\bar{r})=\frac{\pi^{\alpha/2}}{\Gamma(\alpha/2)} \bar{r}^{\alpha-1}, \\
\nabla_\alpha\bar{V}&=\frac{s}{c(\alpha,\bar{r})}\frac{\partial\bar{V}}{\partial\bar{r}},
\end{split}
\end{align}
Here, we have $q = \left((\alpha-1)-\sin(\phi)^2 \right) (\cos(\phi))^{2(1-\alpha)} +    \left((\alpha-1)-\cos(\phi)^2 \right) (\sin(\phi))^{2(1-\alpha)}$,
$p= \cos(\phi)^2(\cos(\phi))^{2(1-\alpha)}+\sin(\phi)^2(\sin(\phi))^{2(1-\alpha)}$,
$s= \cos(\phi)(\cos(\phi))^{1-\alpha}+\sin(\phi)(\sin(\phi))^{1-\alpha}$, where the $\phi$ is the azimuthal angle.
By comparing to the prior results at $\alpha = 1$ case (see Fig below), we find that $\phi = 0$ shows the best agreement. 
At $\phi = 0$, we have $q = \alpha - 1$, $p$ = 1 and $s$ = 1.
In doing so, the solutions of the electric field $\bar{E}= |-\nabla_{\alpha} \bar{V}(\bar{r})|$ and the electric potential $\bar{V}(\bar{r})$ are, respectively, 
$\bar{V}(\bar{r}) = \frac{\bar{r}^{\frac{p+q}{p}}-1}{\bar{b}^{\frac{p+q}{p}}-1}$, and   
$\bar{E}= \frac{\pi ^{-\frac{\alpha }{2}} s \Gamma \left(\frac{\alpha }{2}\right) (p+q) \bar{r}^{\frac{p+q}{p}-\alpha }}{p \left(\bar{b}^{\frac{p+q}{p}}-1\right)}$,
where $\bar{b}= b/a$, $q = \alpha - 1$, $p$ = 1 and $s$ = 1.
The normalized charge $\bar{Q}$ is determined via the surface charge density $\bar{\rho_s}({\bar{r}=\bar{b}})=  \epsilon_r \bar{E}$, which gives
\begin{align}
 \bar{Q}  &= \bar{\rho_s} \bar{r} \int_0^{2\pi} \, d{\phi} \int_0^{\bar{L}} d\bar{z}.
\end{align}
 Finally, the normalized capacitance $\bar{C}(\alpha) = C(\alpha)/C_0$ for a fractal-based porous cylindrical capacitor as a function of $\alpha$ is
\begin{align}
\bar{C}(\alpha)&= \frac{2 \pi ^{1-\frac{\alpha }{2}} s \bar{L} \Gamma \left(\frac{\alpha }{2}\right) (p+q) \epsilon _r \bar{b}^{-\alpha +\frac{p+q}{p}+1}}{p \left(\bar{b}^{\frac{p+q}{p}}-1\right)}.  
\end{align}
Note in the derivation above, the normalized parameters are 
$\bar{E}= E/E_0$, $\bar{V} = V/V_0$, $\bar{r} = r/a$, $\bar{z}=z/a$, $\bar{\rho_s}=\rho_s/\rho_0$, $\bar{Q} = Q/Q_0$, 
$\bar{L}=L/a$, and $\bar{C}= C/C_0$, with normalization constants $E_0 =V_0/a$, $\rho_0 =\epsilon_0 E_0$, $Q_0= a L \rho_0$ and $C_0 = Q_0/V_0$.

\begin{figure}[ht]
    \centering
    \includegraphics[width=12cm, height=5cm]{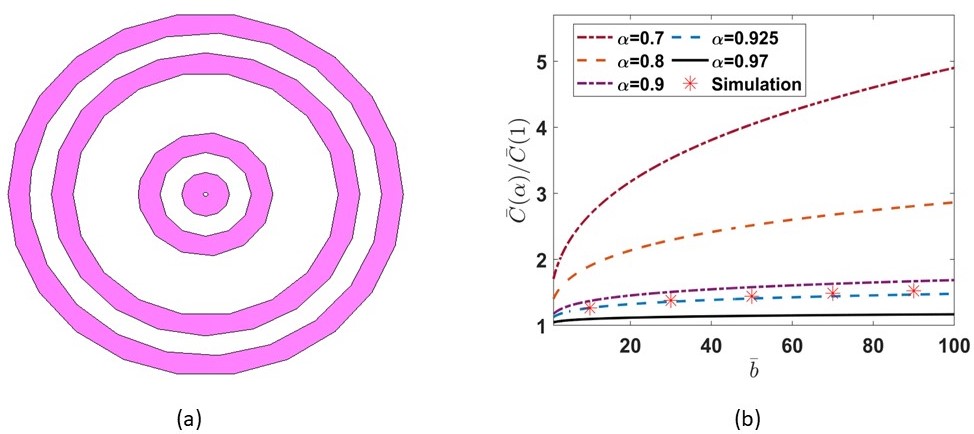}
    \caption{(a) Cross-sectional view of cantor-shaped Mica cylinder (second iteration) with a removal factor of 1/3, which has $\alpha=0.925$ (from box counting method). The pink region is dielectric and the white region is air. (b)  The enhancement of the cylindrical capacitance over the $\alpha$ = 1 limit as a function of $\bar{b}$ for different $\alpha$. The asterisks (at $\alpha=0.925$) are the CST simulation results for comparison.}
    \label{fig:Capacitance.jpg}
\end{figure}

To verify Eq. (29), we create a cylindrical cantor-like (in radial direction) fractal structure with a removal factor of $1/3$ using CST studio suite (see Fig \ref{fig:Capacitance.jpg}(a)).
The pink and white regions are dielectric ($\epsilon_r=6$) and air space, respectively.
Using box-counting method, it has an overall  fractional value of $\alpha=(1.776+1)/3 = 0.925$, where $1.776$ is the fractional value of the 2D cross section of the cylindrical capacitor (Fig. \ref{fig:Capacitance.jpg}a). 
In Fig \ref{fig:Capacitance.jpg}(b), we plot the ratio of $\bar{C}(\alpha)$ to $\bar{C}(1)$ as a function of $\bar{b}$ = 1 to 100 for various $\alpha$ = 0.7, 0.8, 0.9, 0.925, and 0.97.
It shows enhancement as the porosity of a cylinder increases (smaller $\alpha<1$).
At $\alpha$ = 0.925, good agreement is observed between the calculated and simulated results. 


\subsection{Cylindrical MG law for a trap-free solid}
  
 In this section, the 1D MG law for a porous cylindrical diode is derived using the fractional capacitance obtained above.
 When an electron transits the cylindrical gap between anode and cathode, the average normalized electron's velocity $<\bar{\nu}>$ is calculated by
\begin{align}
<\bar{\nu}> &= \frac{\int_1^{\bar{b}} \bar{\nu}(\bar{r})d\bar{r}}{\int_1^{\bar{b}} d\bar{r}^\alpha}= \dfrac{ \alpha \Gamma(\alpha/2)\bar{V}}{ \pi^{\alpha/2}(\bar b^{\alpha }-1)}
\end{align}
where $\bar{\nu}=\nu/\nu_0 $ with $\nu_0 = \mu V_0/a$ and the corresponding average transit time is $<\bar{\tau}> = \int_1^{\bar{b}} d\bar{r}^\alpha / \bar{<\nu>}$.
The normalized current line density is defined as
$\bar{I} =\frac{I/L}{I_0/a}= \dfrac{\bar{Q}}{<\bar{\tau}>}$, and we have
\begin{align}
\begin{split}
\bar{I} &= \frac{2 \pi ^{1-\frac{3 \alpha }{2}} \alpha ^2  s \bar{L} \epsilon_r  \Gamma \left(\frac{\alpha }{2}\right)^3 (p+q) \bar{V}^2\bar{b}^{-\alpha +\frac{q}{p}+2}}{p \left(\bar{b}^{\alpha }-1\right)^2 \left(\bar{b}^{\frac{p+q}{p}}-1\right)}=\frac{2 \pi ^{1-\frac{3 \alpha }{2}} \alpha ^3 \bar{b} \bar{L} \epsilon  \Gamma \left(\frac{\alpha }{2}\right)^3\bar{V}^2}{\left(\bar{b}^{\alpha }-1\right)^3}
\end{split}
\end{align}
where all the constants are $q = \alpha - 1$, $p$ = 1 and $s$ = 1 as shown above.
The normalized time scale and line current density are  $\tau_0 =\mu V_0/a^2$ and $I_0=Q_0 \tau_0^{-1}$. 
Considering the current continuity condition, we determine the electric potential as 
$\bar{V}(\bar{r})=\frac{(\bar{r}^\alpha -1)^{3/2} }{ (\bar{b}^\alpha -1)^{3/2}}$. 
Using this expression, we can obtain the electron velocity and charge density at the anode $(\bar{r} = \bar{b})$ by solving $\nabla_\alpha \bar{V}$ and  $\epsilon_r \times \nabla^2_{\alpha}\bar{V}$ at $\phi=0$ defined in Eq. (27), which are 
$\bar{\nu}(\bar{r}=\bar{b})=\frac{3}{2}\frac{\alpha \Gamma(\alpha/2)}{\pi^{\alpha/2}(\bar{b}^\alpha-1)}$, and 
$\bar{\rho}(\bar{r}=\bar{b})=\frac{3}{4}\frac{  \alpha ^2 \epsilon_r  \Gamma \left(\frac{\alpha }{2}\right)^2}{\pi ^{\alpha } \left(\bar{b}^{\alpha }-1\right)^2}$.
The cylindrical MG law for a porous solid is determined by  $\bar{I}(\alpha)= 2\pi\bar{b}\bar{\nu}(\bar{r}=\bar{b})\bar{\rho}(\bar{r}=\bar{b})$:
\begin{align} \label{eq:frac_cyl_MG}
\begin{split}
\bar{I}(\alpha)&= \frac{9 \pi^{1-\frac{3\alpha}{2}}\alpha^3 \epsilon_r \bar{b} \Gamma(\alpha/2)^3}{4(\bar{b}^\alpha-1)^3}. 
\end{split}
\end{align}


\subsection{Cylindrical MH law for a trap-limited solid}
To extend the cylindrical model for a trap-filled porous solid, we
first, suppose that the injected electrons will fill up the traps bounded given by
\begin{align}
\bar{ Q_t} &=   \bar{\rho_t} \int_0^{2\pi} d\phi \int_{1}^{\bar{b}} \bar{r} d\bar{r}^\alpha  = \bar{C}\bar{V}
\end{align}
Using Eq. (18), the line charge density of the free electrons is
\begin{align}
\bar{Q_f} &=\int_0^{2\pi} d \phi \int_{1}^{\bar{b}} \bar{r}  d \bar{r}^\alpha \quad \bar{e} \bar{N_c} \Big[\dfrac{\bar{C}\bar{V}}{ \bar{e}\bar{N_t }  \int_0^{2\pi} d\phi \int_{1}^{\bar{b}} \bar{r}  d\bar{r}^\alpha } \Big]^l.  
\end{align}
Using this $\bar{Q_f}$ and Eq. (30), the  current line density can be calculated as
$\bar{I} = \bar{Q_f} \times <\bar{\nu}> / \int_{1}^{\bar{b}} d \bar{r}^\alpha$:
\begin{align}
\begin{split}
\bar{I}&=\frac{2^{l+1} \pi ^{1-\frac{\alpha }{2}} \alpha ^2  \Gamma \left(\frac{\alpha }{2}\right) \left(\bar{b}^{\alpha +1}-1\right)\bar{V}}{(\alpha +1) \left(\bar{b}^{\alpha }-1\right)^2}  \left(\frac{\epsilon_r\pi ^{1-\alpha } \alpha  (\alpha +1)   \bar{b} \bar{L} \bar{V} \Gamma \left(\frac{\alpha }{2}\right)^2}{\left(\bar{b}^{\alpha }-1\right) \left(\bar{b}^{\alpha +1}-1\right)}\right)^l.
\end{split}
\end{align}
The electrical potential is in the form of
\begin{align}
\bar{V}(\bar{r}) &= \frac{\left(\bar{r}^{\alpha }-1\right)^{\frac{2+l}{l+1}} \left(\bar{r}^{\alpha +1}-1\right)^{\frac{l-1}{l+1}} }{\left(\bar{b}^{\alpha }-1\right)^{\frac{2+l}{l+1}} \left(\bar{b}^{\alpha +1}-1\right)^{\frac{l-1}{l+1}} },    
\end{align}
and the MH law for a cylindrical porous solid becomes
\begin{align} \label{eq:frac_cyl_MH}
\bar{I}(\alpha)&= 2 \pi \bar{b} \bar{\nu}(\bar{b}) \bar{\rho_f}(\bar{b}).  
\end{align}
Here, $\bar{\nu}(\bar{b})$ and $\bar{\rho_f}(\bar{b})$ is respectively, the electron's velocity and free electron density at the anode, which are 
\begin{align}
 \bar{\nu}(\bar{b}) &= \frac{ \Gamma \left(\frac{\alpha }{2}\right) \{(\alpha +2 \alpha  l+l-1) \bar{b}^{\alpha +1}-(\alpha +1) (l-1) \bar{b}-\alpha  (l+2)\}}{\pi ^{\frac{\alpha }{2}}(l+1) \left(\bar{b}^{\alpha }-1\right) \left(\bar{b}^{\alpha +1}-1\right)}, \\
  \bar{\rho_f}(\bar{b}) &=\bar{e}\bar{N_c} \left(\frac{\bar{\rho_t}(\bar{b})}{\bar{e}\bar{N_t}}    \right)^l, \\
  \bar{\rho_t}(\bar{b}) &= \frac{1}{(l+1)^2 \left(\bar{b}^{\alpha }-1\right)^2 \left(\bar{b}^{\alpha +1}-1\right)^2} \Big[\pi ^{-\alpha } \Gamma \left(\frac{\alpha }{2}\right)^2 \epsilon _r \bar{b}^{-\alpha } \{(5 \alpha -\left(2 \alpha ^2+3 \alpha +1\right) l^2-\nonumber \\& \quad{}2 \alpha  (2 \alpha +1) l+1) \bar{b}^{2 \alpha +1}+\left(-\alpha +\left(2 \alpha ^2+3 \alpha +1\right) l^2+\left(\alpha ^2-2 \alpha -2\right) l+1\right) \bar{b}^{3 \alpha +2}-\nonumber \\& \quad{}(\alpha +1) \left(l^2-1\right) \bar{b}+\alpha ^2 (l+2) \bar{b}^{\alpha }+2 (\alpha +1) (l-1) (2 \alpha +\alpha  l+l+1) \bar{b}^{\alpha +1}+\nonumber \\& \quad{}(\alpha +1) (l-1) (-2 \alpha +l-1) \bar{b}^{\alpha +2}-2 (\alpha +1) (l-1) (\alpha  l+l-1) \bar{b}^{2 \alpha +2}\} \Big].
\end{align}

\begin{figure}[ht]
    \centering
    \subfigure[] { \includegraphics[width=8cm, height=5cm]{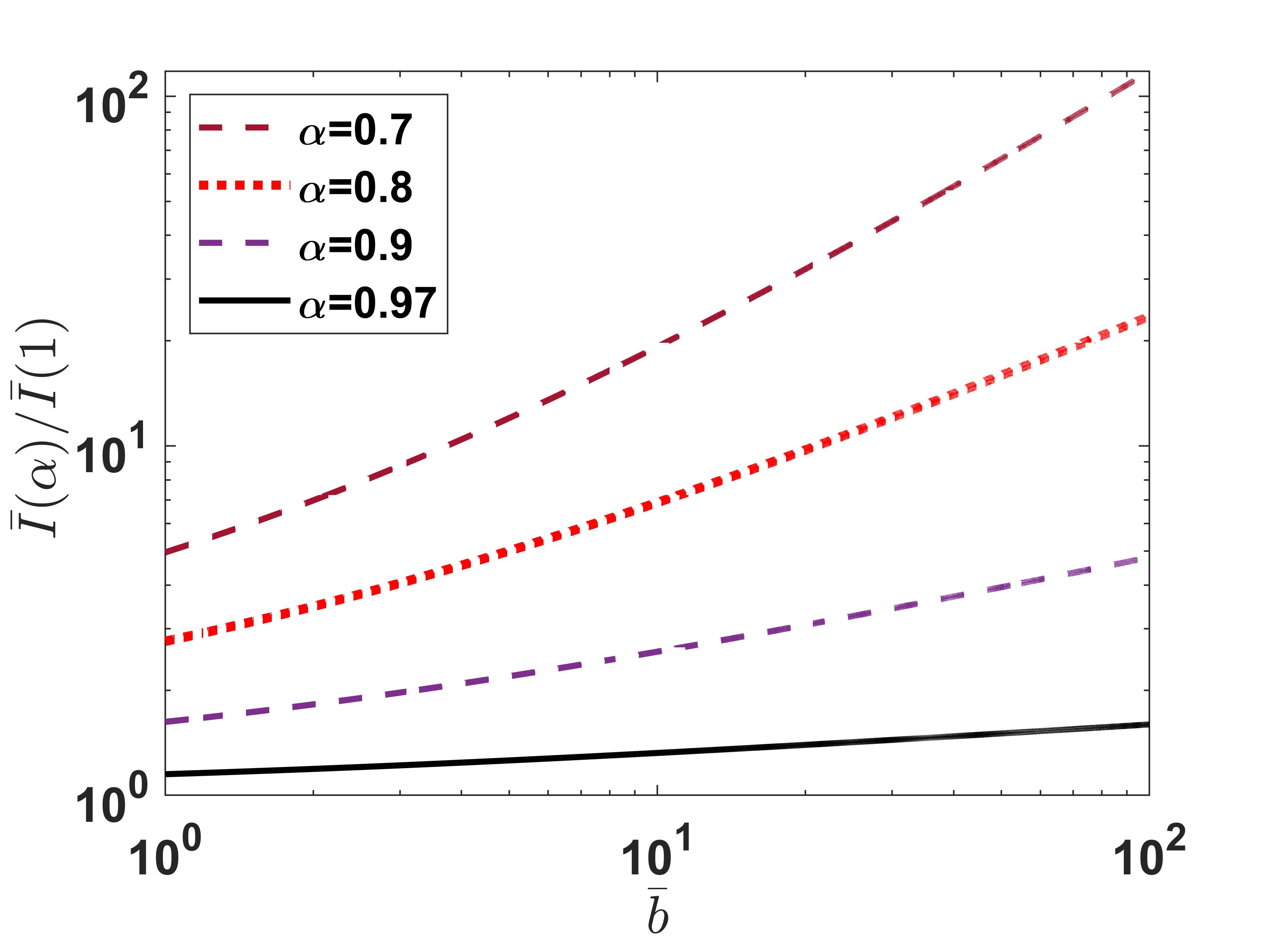}}
    \subfigure[] { \includegraphics[width=8cm, height=5cm]{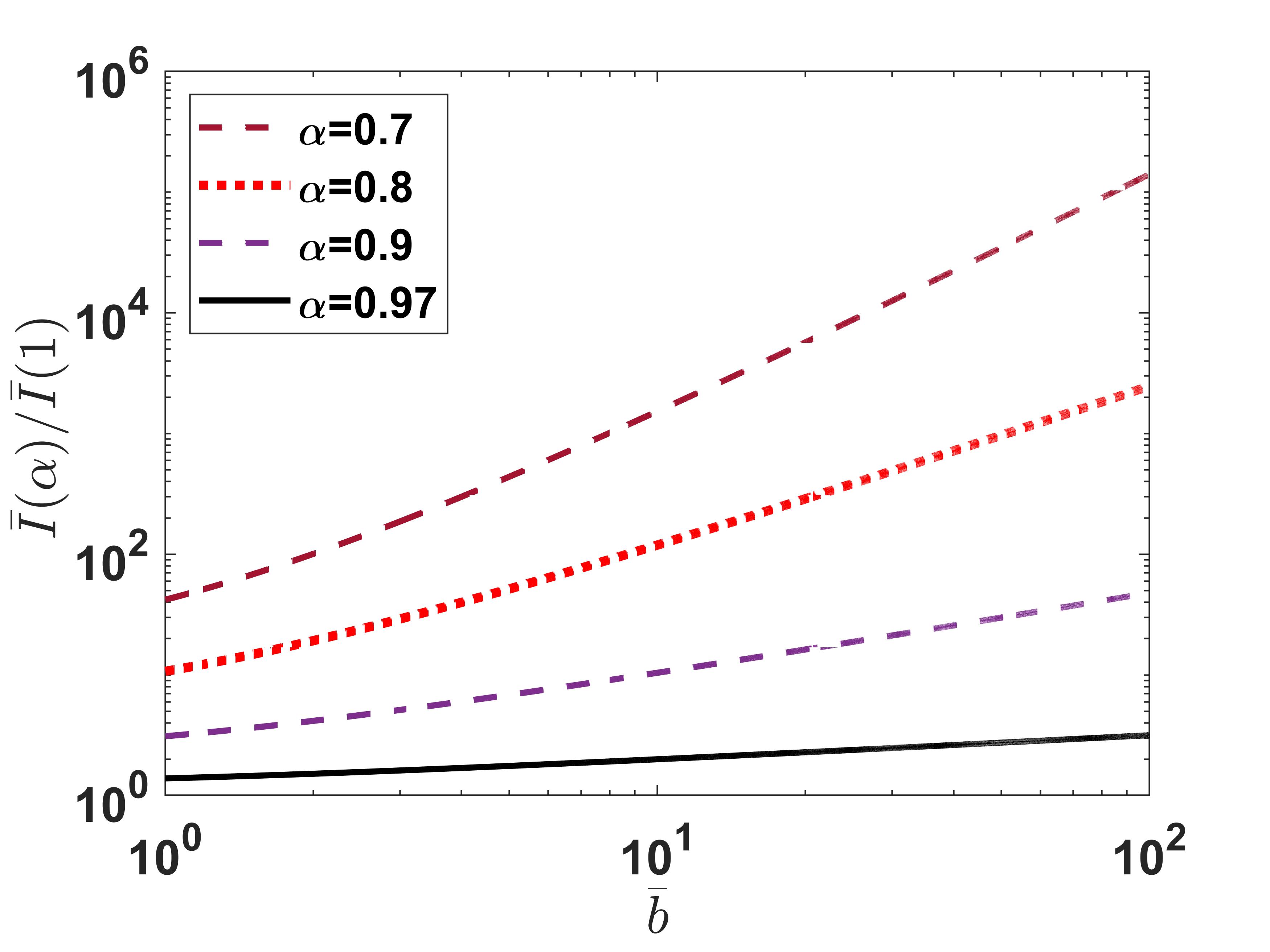}}\\
     \subfigure[] { \includegraphics[width=8cm, height=5cm]{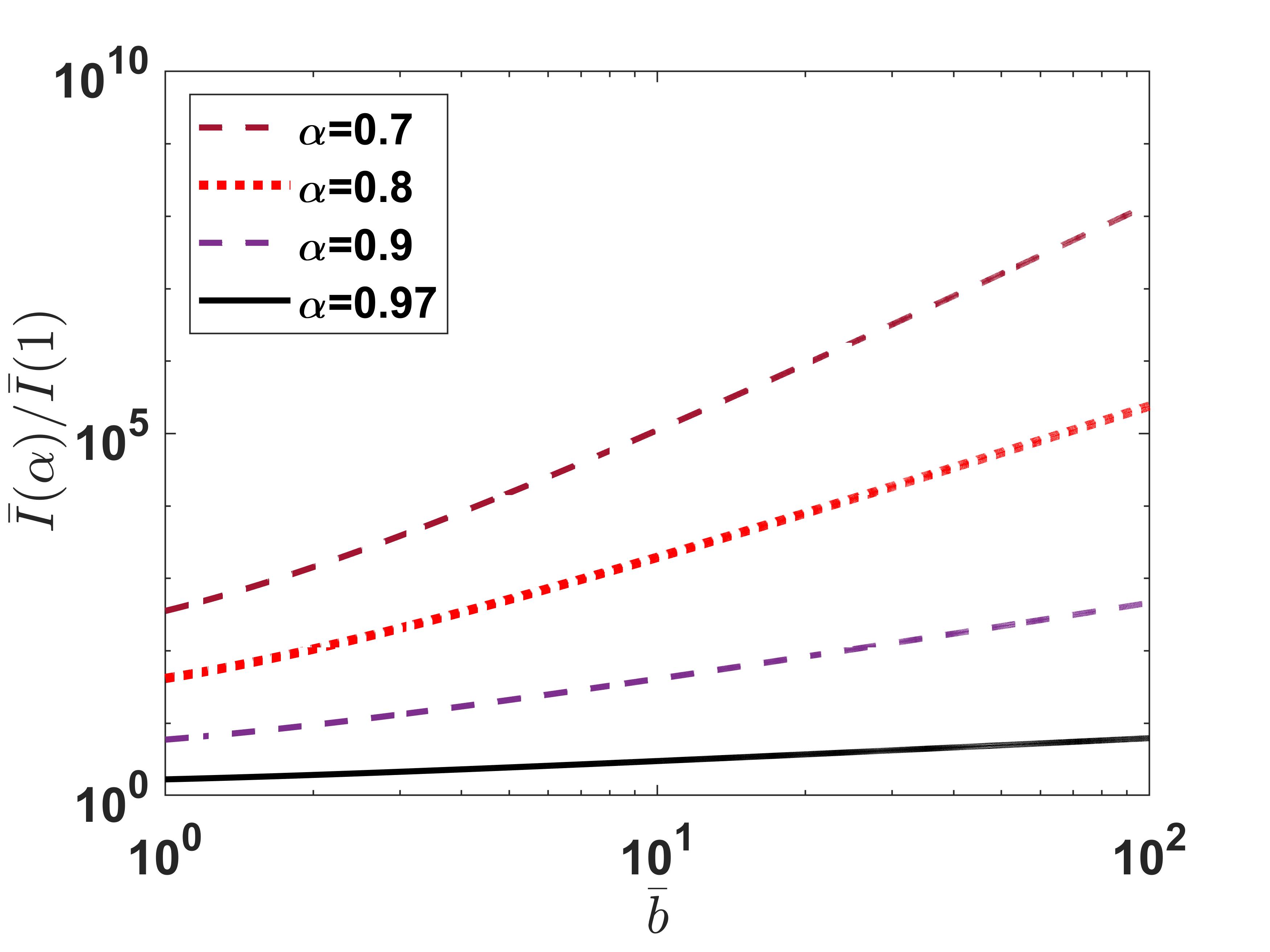}}
      \subfigure[] { \includegraphics[width=8cm, height=5cm]{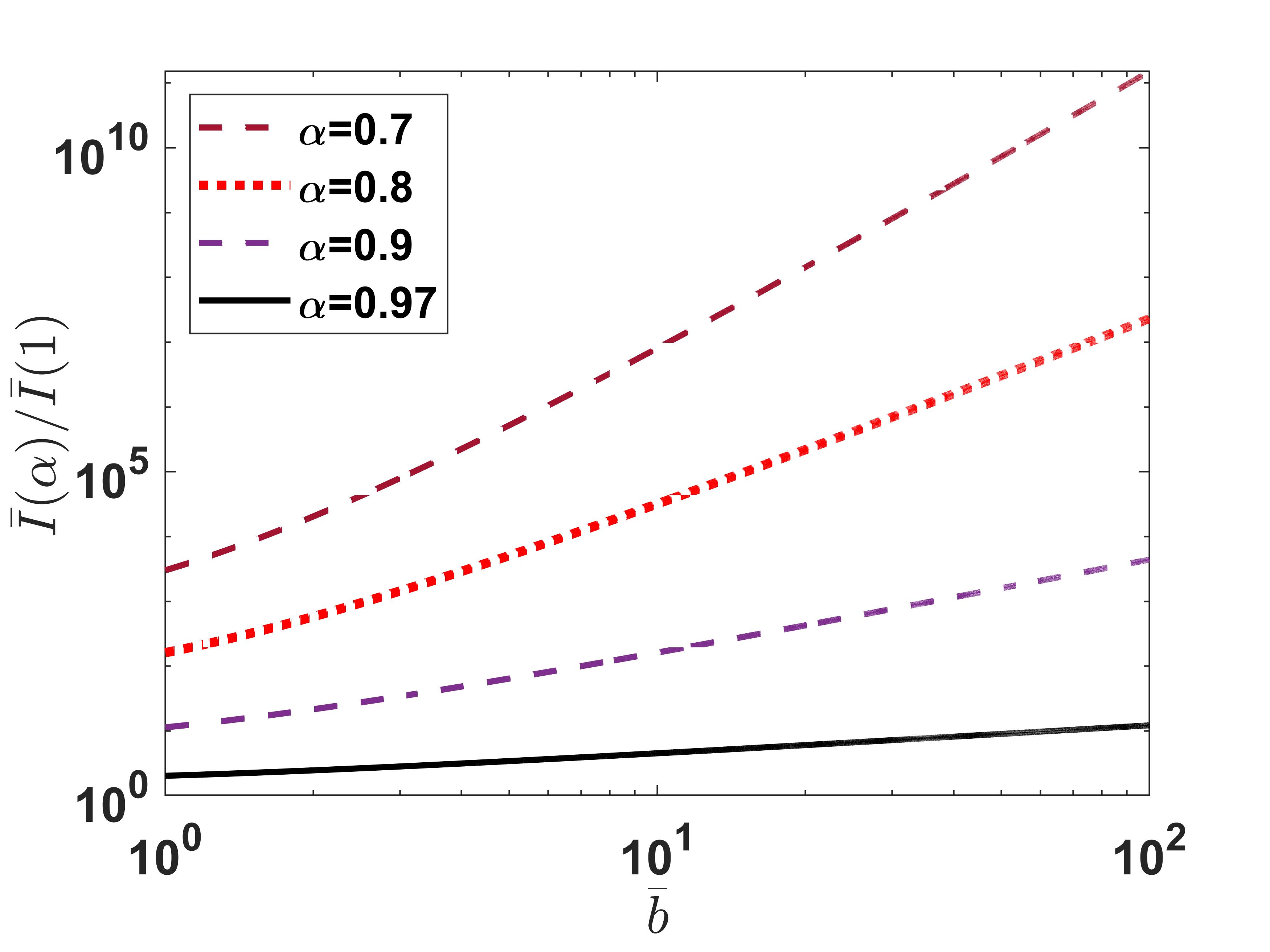}}
    \caption{Enhancement of the cylindrical MH law with $\alpha <$ 1 over the limit at $\alpha$ = 1 as a function of $\bar{b}$ for (a) $l$=1 [MG law], (b) $l$=3, (c) $l$=5, and (d) $l$=7.}
    \label{fig: alpha bar_b}
\end{figure}

In Fig. \ref{fig: alpha bar_b}, we show the the enhancement of the MH law: $\bar{I}(\alpha)$ over its limit at $\alpha =1$ (classical MH law) as a function of $\bar{b}$ = 1 to 100 for $\alpha=$0.7, 0.8, 0.9 and 0.97. 
Four cases of different traps ($l=$1, 3, 5 and 7) are presented to see the effect of trap-limited current.
At $l=$1, we recover the trap-free case (or MG law) as shown in Eq. (\ref{eq:frac_cyl_MG}). 

\begin{figure}[ht]
    \centering
    \subfigure[]{\includegraphics[width=8cm, height=5cm]{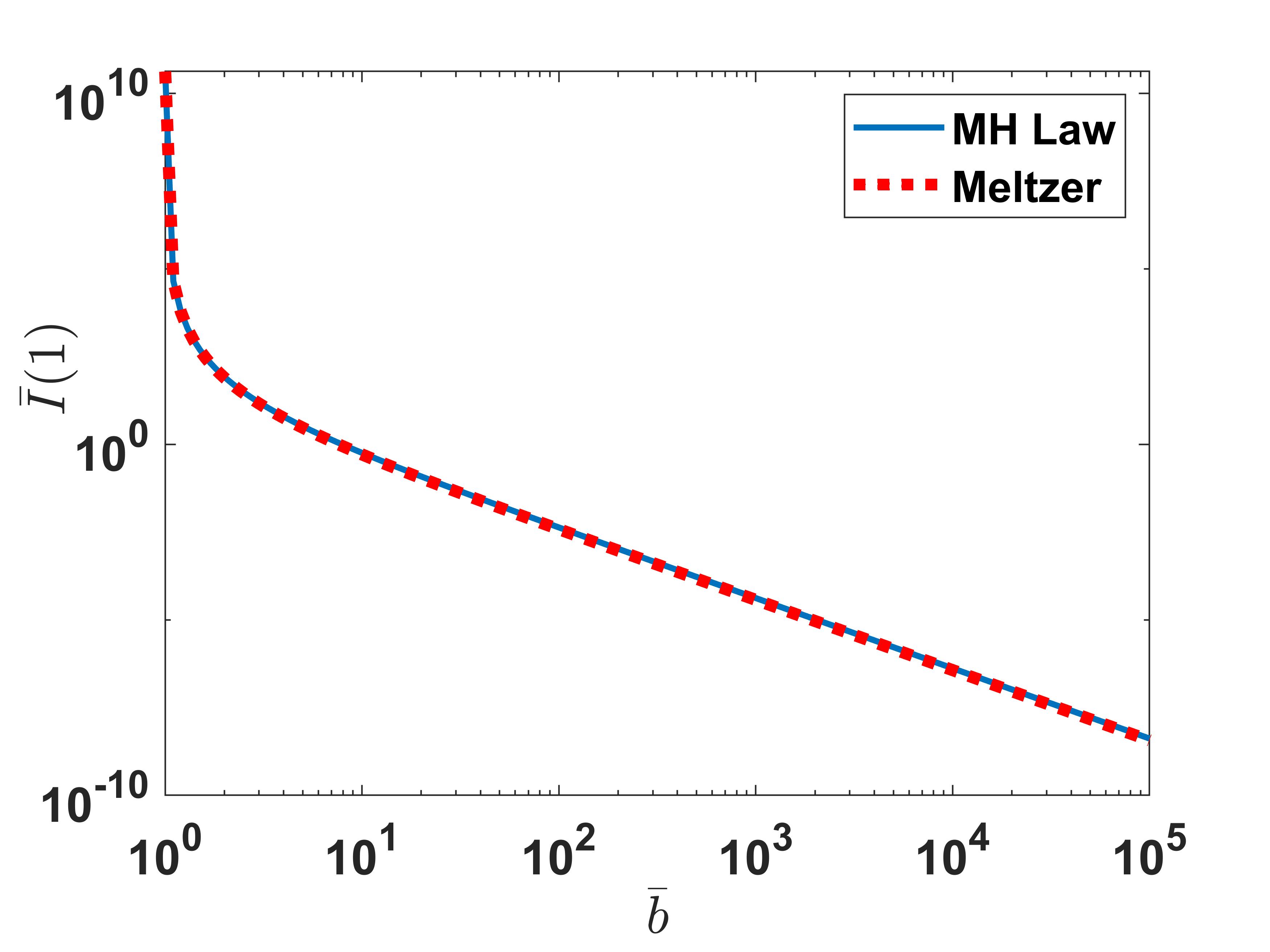}}\subfigure[]{\includegraphics[width=8cm, height=5cm]{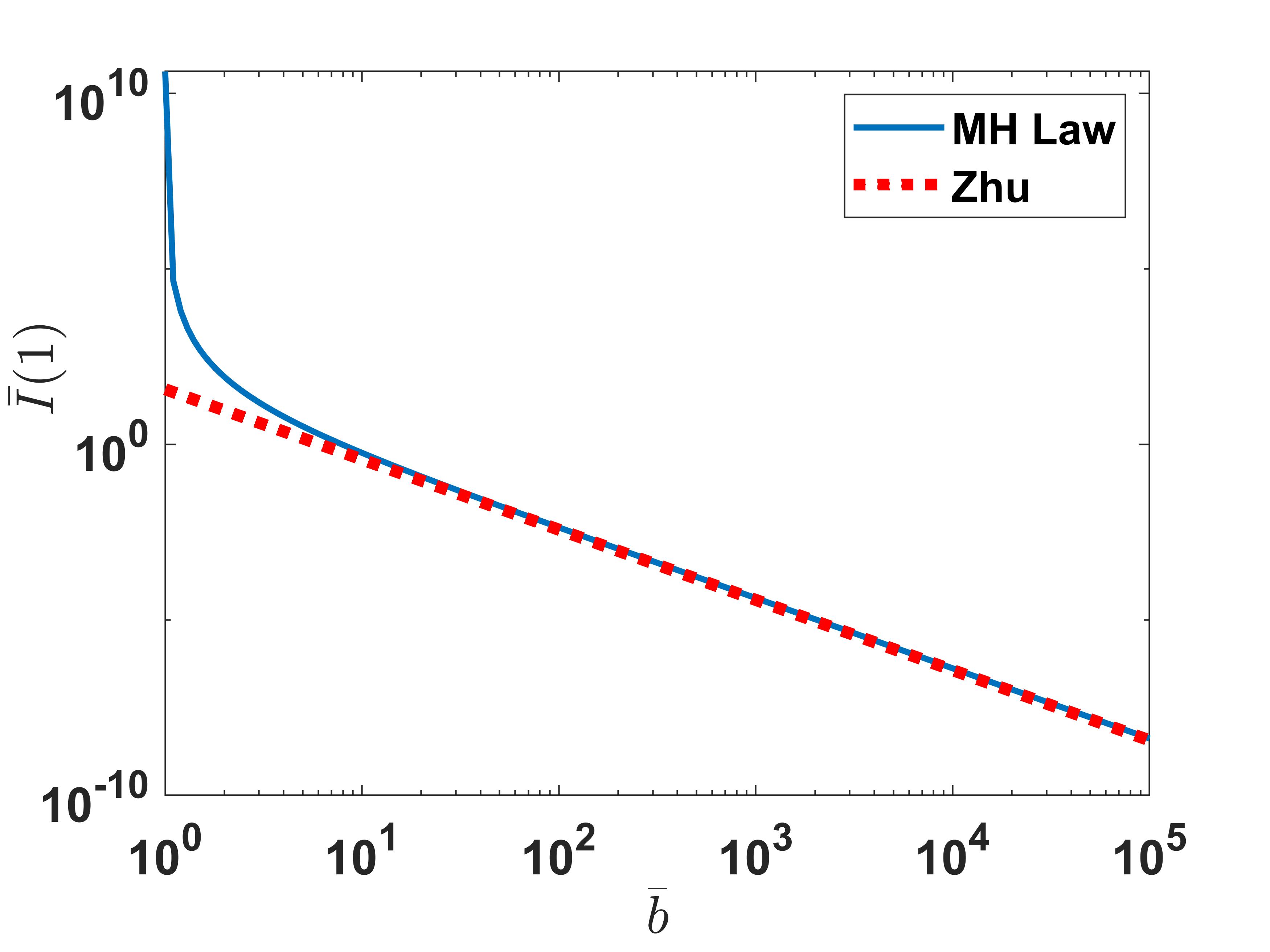}}
    \caption{Comparison of the derived analytical MH law at $\alpha=1$ and $l=1$ from Eq. (\ref{eq:frac_cyl_MH}) with two prior works: (a) by Meltzer \cite{meltzer1960space} and (b) by Zhu \cite{zhu2011analytical} at $\bar{b} \rightarrow \infty$.}
    \label{fig:enter-label}
\end{figure}

Due to the absence of MH law for cylindrical diode without porosity effects (perfect solid), we compare our fractional cylindrical MH law at $\alpha =1$ and $l=1$ (or MG law) with a prior result \cite{meltzer1960space} in Fig. \ref{fig:enter-label}(a), which shows good agreements over a wide range of $\bar{b}$ = 1 to 100.
For cylindrical diode with large aspect ratio of $\bar{b}=b/a \rightarrow \infty$, our analytical results also agree with a prior analytical formulation \cite{zhu2011analytical} at $\bar{b} \geq$ 10.
It is important to note that our newly derived analytical formulation of cylindrical MH law for a porous solid [Eqs. (37) to (40)] is applicable for any $l$, $\alpha$, and $\bar{b}$.


\section{Conclusion}
In this work, we have presented a cylindrical space charge limited current (SCLC) model or MH law for a porous trap-limited dielectric characterized by $\alpha \leq$ 1. 
The analytical model is derived for the first time by using different formulations involving capacitance and transit time models.
The results are verified in recovering the planar case and at $\alpha$ = 1 limit (perfect solid without porosity), such as Mott-Gurney (MG) law and Mark-Helfrich (MH) law for trap-free and trap-limited case respectively.
In the applications of SCLC transport in organic electronics and dielectric breakdown, where the spatial porosity of the material is ubiquitous, this novel model can be a valuable tool for characterizing the porosity by fitting our analytical model to the current-voltage measurements.

\section*{Acknowledgement}
This work is supported by USA ONRG grant N62909-19-1-2047. 
SK acknowledges the support of MOE PhD scholarship.
CYK acknowledges the support of NRF2021-QEP2-02-P03.

\section*{AUTHOR DECLARATIONS}

\section*{Conflict of Interest}
The authors have no conflicts to disclose.

\section*{Data Availability}
The data that support the findings of this study are available from the corresponding author upon reasonable request.

\bibliography{mybib}

\end{document}